# Determination of strain rate in modelling belt-flesh-pelvis interaction during frontal crashes

Zhaonan Sun, Bronislaw D. Gepner, Sang-Hyun Lee, Joshua Rigby, Neil Singh, Jason R. Kerrigan

## I. INTRODUCTION

Obesity is associated with higher fatality risk and altered distribution of occupant injuries in motor vehicle crashes (MVCs) partially because of the increased depth of abdominal soft tissue, which results in limited and/or delayed engagement of the lap belt with the pelvis and increases the risk of pelvis "submarining" under the lap belt exposing occupant's abdomen to belt loading [1-4]. Highly automated vehicles (HAVs) may increase the range of seating positions chosen by vehicle occupants, potentially including an increased prevalence of reclined riding postures. These postures could lead to altered lap belt placement and increased posterior pelvis rotation, potentially increasing the risk of submarining for occupants of all anthropometries, amplifying the challenge of restraining occupants and preventing submarining during MVCs [5-6].

Finite element human body models (FE-HBMs) are valuable tools to simulate occupant response and facilitate the understanding of belt-flesh-pelvis interaction during MVCs. To better facilitate the development of occupant safety countermeasures, biofidelic modeling of the belt-flesh-pelvis interaction during MVCs using FE-HBMs is crucial. Specifically, one critical method to improve biofidelity is to understand the constitutive characteristics of abdominal subcutaneous adipose tissue (SAT) in the belt-flesh-pelvis loading transfer process. However, not only is there a lack of SAT constitutive models applicable to impact biomechanics modelling, no direct measurement for strains and strain rates at the tissue level is available [7]. While estimation was performed to facilitate viscoelastic modeling of human SAT using existing data, the process of the estimation remains unclear in the literature. Therefore, to provide guidelines for future material characterization studies of human SAT, the aim of this study is to estimate the strain levels and strain rates in human SAT during MVCs using FE-HBMs and to analyze the range of applicability of existing material testing data of human SAT to FE-HBMs designed to simulate MVCs.

## II. METHODS

Three types of finite element simulations with different loading and boundary conditions were performed based on the literature and post-mortem human subject (PMHS) tests, with three different FE-HBMs (Figure 1).

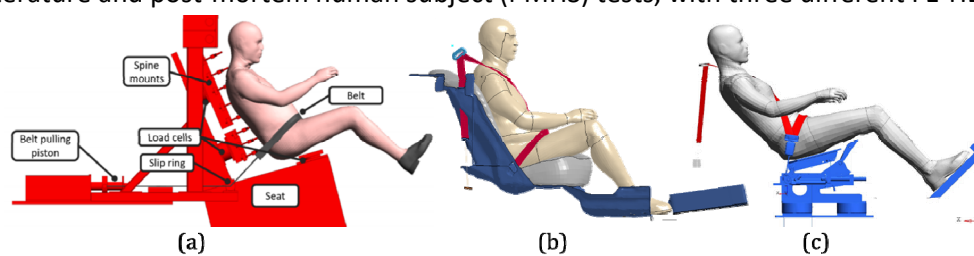

Figure 1. Three types of finite element simulations performed (a) Belt pull simulation; (b) Rear seat simulation; (c) Reclined seatback simulation.

*Belt pull simulation*

The belt pull simulation was based on the belt pull test previously performed at the University of Virginia Center for Applied Biomechanics (UVa-CAB) [2,8]. The FE model of the test fixtures used in the belt pull tests developed previously was used in this simulation. The restraint system includes a 2D lap belt positioned on the occupant, connecting to the 1D belt routed through several sliprings. A belt pulling piston was modelled to perform the belt pulling task based on the force time history, recorded from the experiment. The 30 years old

Z. Sun* is a PhD student (zs2re@virginia.edu, 434-297-8084), B. Gepner and S. Lee are research scientists, and J. R. Kerrigan is an associate professor in Mechanical and Aerospace Engineering at the Center for Applied Biomechanics at the University of Virginia. J. Rigby and N. Singh are undergraduate students in the school of engineering at the University of Virginia.

GHBMC obese model with a height of 1,750 mm and a BMI of 30 kg/m$^2$ (GHBMC-30O) was positioned matching the pelvis and spine position from the PMHS test [2,8].

*Rear seat simulation*

The rear seat simulation was based on PMHS tests performed previously at UVa-CAB [3, 9]. To accurately define the boundary conditions, a previously developed FE model of the test buck was used [10]. The restraint system was modelled based on [10] and included a retractor with pre-tensioner and a force limiter. The 48 km/h acceleration pulse recorded during the test was used as the input. The 70 years old GHBMC with the height of 1,880 mm and BMI of 35 kg/m$^2$ (GHBMC-35O) was positioned in the seat to match the corresponding PMHS. Specifically, the femur angle, tibia angle, H-point, torso angle and arm positions were matched through positioning simulations [2].

*Reclined seatback simulation*

The simulation environment was based on the recently completed reclined, full body, PMHS test series [11]. A previously developed semi-rigid simplified seat, a 50-degree torso recline angle, and a prototype seatback integrated 3-point restraint system were used [6]. Specifically, a 3-point belt equipped with dual lap-belt pretensioners and a shoulder-belt retractor pretensioner was implemented, which was designed to increase engagement of the pelvis, reducing the risk of submarining for reclined occupants. Also, no knee bolster was used. A 50 km/h full frontal rigid barrier pulse with a peak acceleration of 35g and a delta-v of 51 km/h [11-13] was used as the input. A 50$^{th}$ percentile male FE-HBM model, the GHBMC simplified occupant model (GHBMC-S) version 1.8.4, was used in this simulation. The GHBMC-S was positioned to match the approximated PMHS positions using initial data from the external retroreflective body markers [14].

*Strain and strain rate calculation*

Upon completion of the simulations, element strains in the abdominal flesh part were extracted. Specifically, the infinitesimal strain tensor was obtained, and three strain metrics were derived from the strain tensor. The maximum principal strain (MAPS) and minimum principal strain (MIPS) were defined as the maximum and minimum numerical value among the three principal strains, which correspond to either tensile or compressive loading of the elements. Shear strain was obtained by extracting the off-diagonal element in the infinitesimal strain tensor which correspond to element shearing in the sagittal plane of the human body. The peak strains of each element were obtained by taking the peak strain value in the simulation time history based on these three strain metrics. Among these individual peak strain values, the 95$^{th}$ percentile value was obtained and presented as overall peak strains to avoid numerical artifacts contributed by heavily distorted elements. Strain rate at each element was obtained by dividing the difference in element strain across two consecutive timesteps by the timestep length. Then, the peak strain rate value for each element was obtained by finding the peak value across the simulation time history. Similarly, the 95$^{th}$ percentile value of peak strain rates among these individual peak strain rates in the simulation was then calculated and presented as peak strain rates.

The above procedure was performed for all three simulations and the average value of the overall peak strain and strain rates was calculated across different simulations. To isolate the loading effect in the area of direct contact with the lap belt, the same procedure was repeated for the elements in the abdominal region under direct belt loading. The obtained values were compared with input conditions from material testing of animal and human SAT in the literature.

### III. INITIAL FINDINGS

In reclined seatback with GHBMC-S, despite the reclined position, the occupant's pelvis was well constrained and no pelvis submarining under the lap belt was observed. Specifically, the lap belt remained anterior to the anterior superior iliac spine (ASIS) throughout the simulation. Large deformation was observed in the abdominal flesh part while the torso was moving forward. In the rear seat simulation with GHBMC-35O, while the forward motion of lower extremities was significant, no pelvis submarining under the lap belt was observed. As a result, the flesh covering the pelvis was loaded to large deformation. In the belt pull simulation, the pelvis of the GHBMC-30O was constrained to match the PMHS test. While the lap belt in the belt pull simulation followed the initial belt trajectory observed in the experiment, it stayed engaged with the flesh covering the pelvis and failed to slide into the abdomen.

Considering all elements in the abdominal flesh part, the largest peak MAPS across all simulations was found to be 0.5 in the GHBMC-S model while the largest peak rate for MAPS was 44/s, observed in the belt pull simulation with the GHBMC-35O model. Both the largest peak MIPS and peak MIPS rate was observed in the reclined sled simulation with the GHBMC-S model. Peak shear strain was found in the rear seat sled simulation while the peak shear strain rate was observed in the belt pull simulation. Overall, the average strain and strain rate MAPS response, for all simulations considering all elements in the abdominal flesh part were 0.46 and 35/s, respectively. The analyzed result using the other two metrics are presented in Table 1 along with detailed information on each FE-HBM. Values in the peak MIPS column were negative due to the sign convention that represents compressive strain in negative numbers. The peak MIPS rate, however, was represented as positive numbers for the purpose of comparison with other metrics.

Table 1. Strain and strain rate from all elements of abdominal flesh part

| Test | Model | Peak MAPS | Peak MAPS rate | Peak MIPS | Peak MIPS rate | Peak shear strain | Peak shear strain rate |
|---|---|---|---|---|---|---|---|
| Belt Pull | GHBMC-35O | 0.42 | 44 | -0.59 | 8 | 0.36 | 37 |
| Rear seat sled | GHBMC-30O | 0.47 | 21 | -0.81 | 7 | 0.42 | 15 |
| Reclined Sled | GHBMC-S | 0.50 | 41 | -0.87 | 20 | 0.22 | 32 |
| Average response | N/A | 0.46 | 35 | -0.76 | 12 | 0.33 | 28 |

When only considering elements in contact with the lap belt, it was found that the highest peak values for both MAPS metrics and shear strain metrics were observed in the belt pull simulation with the GHBMC-35O model. Peak MIPS and MIPS rate were found in the rear seat sled simulation. The average peak strain rates appeared to be similar to the result obtained with all elements in the abdominal flesh part. The detailed information was presented in Table 2.

Table 2. Strain and strain rate from elements in contact with lap belt

| Test | Model | Peak MAPS | Peak MAPS rate | Peak MIPS | Peak MIPS rate | Peak shear strain | Peak shear strain rate |
|---|---|---|---|---|---|---|---|
| Belt pull | GHBMC-35O | 0.53 | 49 | -0.73 | 3 | 0.53 | 42 |
| Rear seat sled | GHBMC-30O | 0.45 | 12 | -0.98 | 26 | 0.05 | 5 |
| Reclined Sled | GHBMC-S | 0.50 | 46 | -0.95 | 11 | 0.26 | 35 |
| Average response | N/A | 0.49 | 36 | -0.89 | 13 | 0.28 | 27 |

## IV. DISCUSSION

While pelvis submarining was observed in the UVA belt-pull test and rear seat sled test with PMHS, different belt-flesh-pelvis interaction patterns were observed in the corresponding simulations. One of the reasons for this discrepancy is that the FE-HBM flesh material model does not represent the constitutive behavior of human SAT [15]. To characterize the desired constitutive behavior and develop a biofidelic flesh model, it is important to understand the deformation from these simulations. Since material characterization and analytical modelling are generally performed using uniaxial or biaxial tensile tests, compression tests and shear tests, MAPS, MIPS and shear strain metrics were chosen to guide those tests effectively. In general, the results showed that human SAT underwent large deformation under high loading rates. While either the average response in the contact area or the whole part shall be used as a guideline for material testing, the variance of strain and strain rates in different simulations showed the necessity of biofidelic material model suitable for large deformation at various strain rates, preferably from 10/s to possibly 50/s. It should be noted that, since truncation was consistently set at 95th percentile when calculating the maxima, the local maximum was sometimes greater than its corresponding global counterpart because distribution of involved elements was altered.

The results indicate deep compression in abdominal elements, which is partially related to the computational modelling technique of the abdominal flesh. In all three FE-HBMs, severe volume change of the flesh was observed since it was modelled as compressible material with a Poisson's ratio of 0.3 whereas it is assumed that human SAT should be modelled as nearly incompressible material, similar to other soft tissues [16]. This desired incompressible material characteristics result in higher compressive stiffness relative to shear stiffness. Therefore, it is assumed that the biofidelic material characteristic of SAT would make the abdomen of FE-HBMs prone to shear deformation and could facilitate the models to replicate the submarining kinematics.

To improve the biofidelity of modeling belt-flesh-pelvis interactions in FE-HBMs, preferably a rate sensitive model applicable at high loading rates under large deformation should be implemented, capturing responses from high rate to quasi-static loading. However, most studies on human SAT were carried out with combinations of either large deformation at low strain rates, or small strain at high strain rates such as rheometer tests [7]. Since the adipose tissue was shown to be nonlinear and rate dependent and very little is known about adipose tissue behavior in the realm of large deformation at high strain rates, the results from these studies cannot be directly applied to FE-HBMs for MVC simulations.

Since the elements in the finite element simulation are under combined loading, it is necessary to perform material testing and characterization of human SAT in different loading modes. Previous research has concluded that human SAT is primarily under compression and shear in MVCs [2]. Also, the difference in material response between compression and shear strain at various strain rates made it difficult for the constitutive model developed from uniaxial tests only or shear tests only to match the responses in both compression and shear simultaneously. Therefore, the constitutive model fitted with data from only one loading mode should be used with caution.

While it is necessary to perform constitutive modelling of human SAT using data at large deformation, high rate and from different loading modes, no existing data set in the literature is available to perform such constitutive modelling for human SAT. Several studies have tested human SAT under various loading modes while none of these studies were aimed at applications similar to abdominal belt loading in MVCs. Human SAT was tested under biaxial tensile and shear tests quasistatically [17]. Although a stress relaxation curve was presented, the applicability of interpolating their data into higher rate loading cases remained unverified. Uniaxial tensile test and stress relaxation test for human SAT were performed at a strain rate up to 0.15%/s [18]. While a linear viscoelastic model was reported, the applicability at higher loading rate remained questionable. Rheology study for shear was performed, reporting the viscosity, elastic modulus, viscous modulus and complex modulus obtained from shear tests up to 20% shear strain [19]. Overall, the applicability of these characterized SAT materials from a single study should be used with caution since the input of the test might not match the loading condition in MVCs. While attempts have been made to fit a constitutive model by pooling existing data from multiple sources with a fully nonlinear viscoelastic model using the Reese and Govindjee formulation [7,20], the data fitted into the model was obtained from porcine SAT instead of human SAT. Also, no available validation data set is available to calibrate the model response for large deformation at high loading rates. Therefore, future study focusing on characterizing the material properties of human SAT under large deformation, at high loading rate and from different loading modes are necessary. to further improve FE-HBMs for simulating occupant responses in MVCs. Existing quasistatic data can also be used in conjunction with future data to calibrate rate sensitive material behavior of human SAT.

While this study aims to shed light into loading conditions of human SAT material testing, one underlying assumption of this work is that calculated strain and strain rates depend on the constitutive model of human SAT and the mesh quality in the FE-HBMs. Therefore, the results should be used as a baseline for a wide range of testing across different rates. Future work should also include re-estimating the amount of deformation and strain rate in the abdominal flesh part after mechanical testing and constitutive modelling of human SAT.

This study demonstrated a new comprehensive approach for estimating the expected strain rates and strain levels for human SAT during MVCs. It was found that the human abdomen was under large deformation at high strain rates. Also, no existing data set is available in the literature for analytical modelling of human SAT in such loading conditions. This effort will inform future investigation of human SAT properties with the application of improved SAT modeling on existing HBMs to improve belt-flesh-pelvis interaction modelling during MVCs. This will eventually enhance the HBMs prediction of submarining injury risk and their application to designing optimized restraint systems.